# Studies of dense nuclear matter at NICA

Peter Senger[1,2]

[1] Facility for Antiproton and Ion Research (FAIR), Darmstadt, Germany
[2] National Research Nuclear University MEPhI, Moscow, Russia

**Abstract**
Laboratory experiments with high-energetic heavy-ion collisions offer the opportunity to explore fundamental properties of nuclear matter, such as the high-density equation-of-state, which governs the structure and dynamics of cosmic objects and phenomena like neutron stars, supernova explosions, and neutron star mergers. A particular goal and challenge of the experiments is to unravel the microscopic degrees-of-freedom of strongly interaction matter at high density, including the search for phase transitions, which may feature a region of phase coexistence and a critical endpoint. As the theory of strong interaction is not able to make firm predictions for the structure and the properties of matter high baryon chemical potentials, the scientific progress in this field is driven by experimental results. The mission of future experiments at FAIR and NICA, which will complement the running experimental programs at GSI, CERN, and RHIC, is to explore new diagnostic probes, which never have been measured before at collision energies, where the highest net-baryon densities will be created. The most promising observables, which are expected to shed light on the nature of high-density QCD matter, comprise the collective flow of identified particles including multi-strange (anti-) hyperons, fluctuations and correlations, lepton pairs, and charmed particles. In the following, the perspectives for experiments in the NICA energy range will be discussed.

## 1. High-density QCD matter in the cosmos and in the laboratory

The exploration of strongly interacting matter under extreme conditions is one of the most exciting research programs of modern nuclear physics. The recent discoveries of neutron star mergers and supermassive neutron stars challenge our knowledge on high-density QCD matter, like its equation-of- state (EOS) and the microscopic degrees-of-freedom. Laboratory experiments with energetic heavy-ion collisions offer the possibility explore the high-density EOS and to investigate new phases of matter, which may feature characteristic structures such as a first-order phase transition with a region of phase coexistence and a critical endpoint. The experimental discovery of these prominent landmarks of the QCD phase diagram would be a major breakthrough in our understanding of the properties of nuclear matter, with fundamental consequences for our knowledge on the structure of neutron stars and the dynamics of neutron star collisions. Equally important is quantitative experimental information on the properties of hadrons in dense matter, which may shed light on chiral symmetry restoration and the origin of hadron masses. Worldwide, substantial efforts at the major heavy-ion accelerators are devoted to the clarification of these fundamental questions, and new dedicated experiments are under construction at future facilities like FAIR in Darmstadt and NICA in Dubna.

The conjectured phases of nuclear matter and their boundaries are illustrated in figure 1 in a 3-D diagram of temperature versus baryon chemical potential and isospin chemical potential [1]. At very low baryon-chemical potentials and high temperatures, the fundamental theory of strong interaction, Quantum Chromo Dynamics (QCD), predicts a smooth crossover transition from hadronic matter to the quark-gluon plasma at a pseudo-critical temperature of about 155 MeV

[2,3], which is more than hundred-thousand times hotter than the core of the sun. Above this temperature, QCD matter dissolves into its constituents, and exists as a mixture of elementary matter and antimatter, similar to the primordial soup in early universe, about a microsecond after the big bang. Matter with equal numbers of particles and antiparticles, similar to the conditions in the early universe, is also produced heavy-ion collision experiments at LHC and top RHIC energies. When analyzing the measured particle yields with statistical hadronization models, a freeze-out temperature of about T = 156 MeV is found, at almost zero chemical potential [4,5]. This temperature coincides with the critical temperature predicted by Lattice QCD calculations for a chiral phase transition as mentioned above.

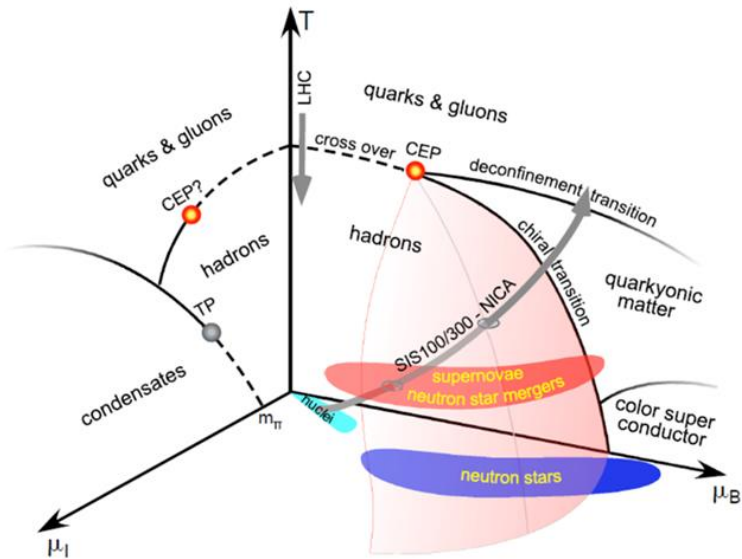

Figure 1: Sketch of the phase diagram for nuclear matter. Taken from [1].

The structure of the QCD phase diagram in the region of large baryon chemical potential is essentially unknown. Lattice QCD calculations are still limited to very low $\mu_B$, and, hence, leave room for effective models, which predict a first-order phase transition with a critical endpoint [6]. None of these structures have been found so far. Moreover, exotic phases of QCD matter have been proposed, such as quarkyonic matter, which can be considered as a Fermi gas of free quarks, with all thermal excitations permanently confined [7]. In addition to these conjectured landmarks, figure 1 indicates also the regions of cosmic matter like neutron stars and neutron star mergers, which are located at large baryon-chemical potentials, low temperatures, and finite isospin chemical potentials.

The existence of a first order phase transition ending in a critical point as indicated in figure 1 is still under debate. For example, following the concept of quark-hadron continuity, it is predicted that quark degrees of freedom emerge gradually with increasing density and partial restoration of chiral symmetry [8]. This scenario is illustrated in figure 2, which depicts a schematic picture of a continuous transition from nuclear to deconfined quark matter with increasing density. For densities below 2 $\rho_0$, the dominant interactions occur via a few meson or quark exchanges and the matter can be described in terms of interacting nucleons. For densities in the range from 2 $\rho_0$ to 4 – 7 $\rho_0$, many-quark exchanges dominate, and the system gradually changes from hadronic to quark matter. For densities beyond 4 – 7 $\rho_0$, nucleons percolate, start to melt and to dissolve into their constituents, which do not longer belong to particular nucleons. A perturbative QCD description is expected to be valid only for extremely high densities, that is, above 100 $\rho_0$.

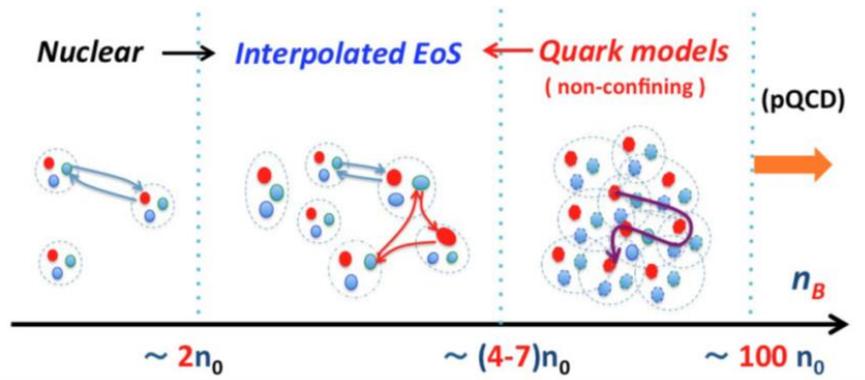

Fig. 2. Sketch of a suggested continuous the transition from nuclear to quark matter with increasing density in units of saturation density (see text). Figure courtesy of T. Kojo [8].

### 1.1 Dense QCD matter in compact stellar objects

Various calculations have been performed which indicate phase transitions from hadronic to quark matter in the core of massive neutron stars, in neutron star mergers or even in core collapse supernovae of very massive stars. Figure 3 depicts the result of calculations based on a non-local 3-flavor Nambu Jona-Lasinio model, which predicts a mixed phase where hadrons including hyperons and quarks coexist in the core of a 2 solar mass neutron star above densities of about 5 $\rho_0$ [9]. In this case, repulsive vector interactions among the quarks are introduced to prevent a softening of the EOS, which otherwise would happen due to the appearance of quarks and hyperons. Above densities of 8 $\rho_0$ pure quark matter is predicted.

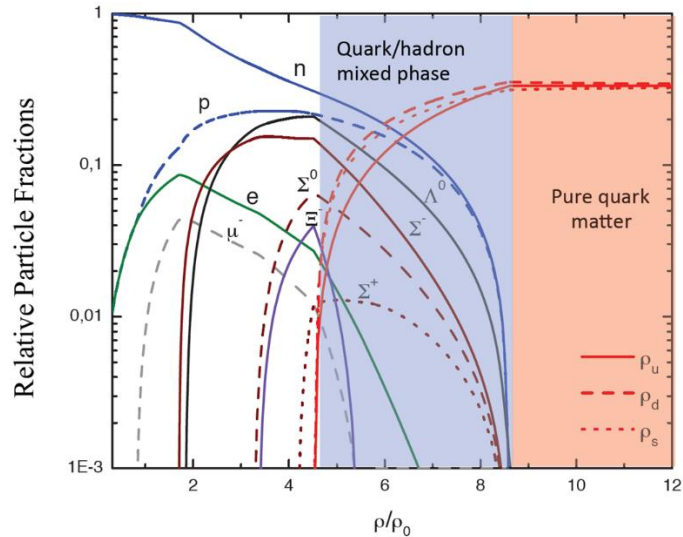

Fig.3: Particle population in a neutron star calculated with a non-local 3-flavor Nambu Jona-Lasinio (n3NJL) model with repulsive vector interactions. The model is able to describe neutron stars with 2 solar masses and radii between 12 and 13 km [9].

Densities up to 4 $\rho_0$ are also expected to be reached also in neutron star mergers as illustrated in figure 4, which depicts a snapshot of the equatorial plane illustrating the evolution of a neutron star merger with a total mass of 2.8 solar masses calculated with a Chiral Mean Field model [10]. The left part of the plot displays the temperature T, while the right part presents the quark fraction $Y_{quark}$. The green lines represent contours of constant baryon density in units of the nuclear saturation density $\rho_0$. The calculation predicts a phase transition to pure quark matter at a density of 4 $\rho_0$ and at a temperature of about 50 MeV. This phase transition occurs shortly before the high-mass neutron star collapses into a black hole.

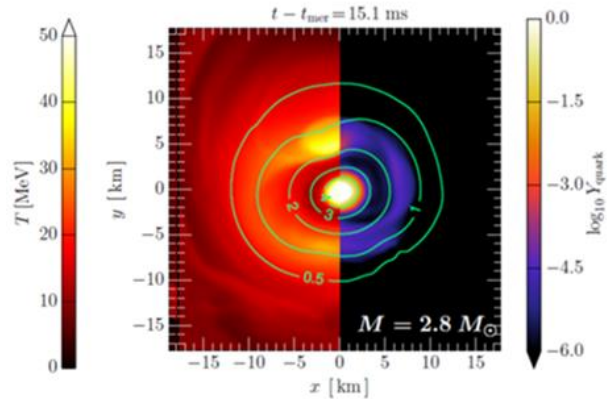

Fig.4: Snapshot of the evolution of two merging neutron stars with a total mass of 2.8 solar masses. The left and the right part of the plot indicates the temperature and the $\log_{10}$ of the quark fraction $Y_{quark}$, respectively. The green lines represent contours of constant net-baryon density in units of $\rho_0$. Adapted from [10].

The density reached in neutron stars and neutron star mergers depends critically on the nuclear matter equation-of-state (EOS). This is illustrated in figure 4, where the mass of neutron stars is plotted versus their central density for various EOS [11]. The recent relativistic Shapiro delay measurement of an extremely massive millisecond pulsar found a mass of 2.14 + 0.10 − 0.09 solar masses [12], which corresponds, according to figure 4, to a central density between 4 and 7 $\rho_0$.

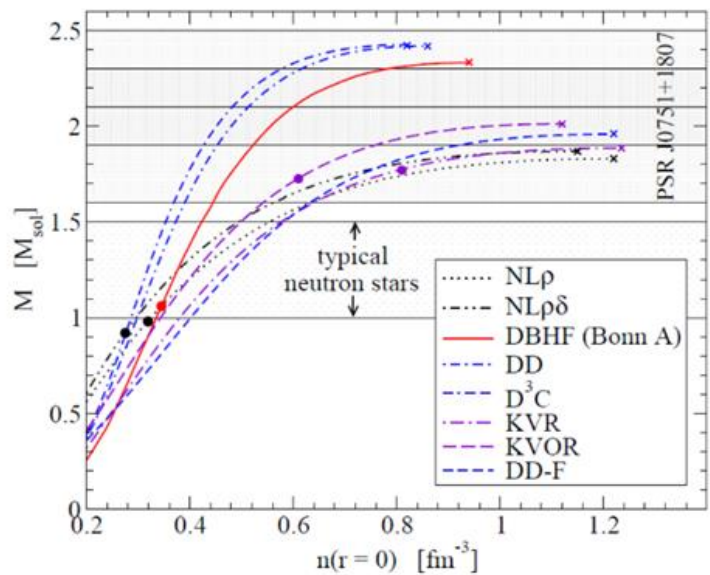

Fig. 4: Mass versus central density of neutron stars calculated from the Tolman–Oppenheimer–Volkoff (TOF) equation using different equation-of-state (EOS) [11].

As illustrated in figure 4 and according to the TOF equation, mass and radius of neutron stars is determined by the high-density EOS. The EOS describes the relation between density, pressure, volume, temperature, energy, and isospin asymmetry. For a constant temperature, the pressure can be written as

$$P = \rho^2 \delta(E/A)/\delta\rho \quad (1)$$

where P represents the pressure, $\rho$ the density, and E/A the energy per nucleon, which is defined as

$$E/A(\rho,\delta) = E/A(\rho,0) + E_{sym}(\rho)\cdot\delta^2 + O(\delta^4) \quad (2)$$

with the symmetry energy $E_{sym}$ and the asymmetry parameter $\delta = (\rho_n-\rho_p)/\rho$. The EOS for symmetric matter has a minimum value at saturation density $\rho_0$, and the curvature around $\rho_0$ is parameterized by the nuclear incompressibility $K_{nm} = 9\rho^2\,\delta^2(E/A)/\delta\rho^2$. Various EOS calculated for symmetric nuclear matter are shown in the left panel of figure 5, while different models for the symmetry energy are depicted in the right panel of figure 5 [13]. The nuclear incompressibility for almost symmetric matter at saturation density has been extracted from measurements of giant monopole resonances of heavy nuclei, and found to be $K_{nm}(\rho_0) = 230\pm10$ MeV, corresponding to a relatively soft EOS [14]. The symmetry energy at saturation density has been derived from nuclear masses and neutron skin measurements, and found to be $E_{sym}(\rho_0) \approx 32$ MeV [15]. Therefore, most of the models for $E_{sym}$ shown in the right panel of figure 5 agree on this value.

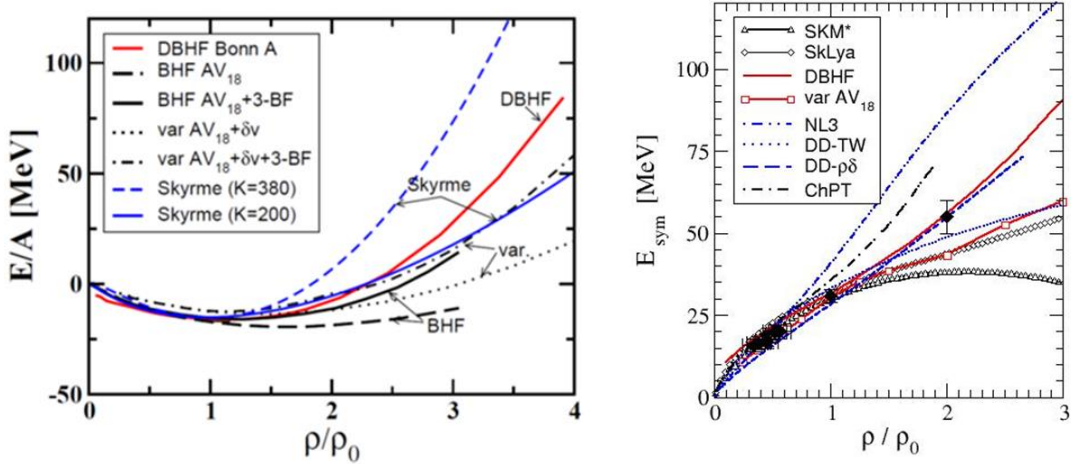

Fig. 5: Left: The isospin-symmetric nuclear matter EOS based on Skyrme forces for different incompressibility compared to the predictions from microscopic ab initio calculations. Right: Symmetry energy for different EOS as function of density [13].

It is worthwhile to note, that the EOS for neutron star matter is defined as the sum of the EOS for symmetric matter and the symmetry energy according to equation 2, but for densities above 4-5 $\rho_0$, where the model extrapolations are highly uncertain. Therefore, various astrophysics experiments try to constrain the EOS at high densities. For example, the Neutron Star Interior Composition Explorer (NICER) detector located at the International Space Station measures time-resolved x-ray spectra emitted from hot spots of neutron stars, in order to extract simultaneously information on their mass and radius [16]. Moreover, constraints on the radii of neutron stars have been derived by multi-messenger observations of neutron star mergers in combination with nuclear theory [17]. Once the mass and radius of a nucleon star has been measured with high accuracy, the high-density EOS can be calculated via the Tolman-Oppenheimer-Volkoff (TOF) equation. Complementary to astrophysical observations, the high-density EOS can be studied in heavy-ion collisions. Laboratory experiments in addition offer the possibility, to explore the microscopic degrees-of-freedom at high density, which is another fundamental property of QCD matter.

## 1.2 Laboratory experiments exploring dense QCD matter

### 1.2.1 Investigations of nuclear matter at high net-baryon densities

Pioneering experiments at the Bevalac in Berkeley in the 1980ies discovered the collective flow of protons, indicating that a new form of equilibrated and dense matter is created in heavy-ion collisions at beam energies around 1A GeV [18]. Moreover, the multiplicity of pions emitted from this fireball was used to extract the EOS of nuclear matter [19]. More quantitative information on the EOS was obtained by the next generation of experiments in the same beam energy range, which have been performed in the 1990ies at GSI. The FOPI collaboration measured the elliptic flow of protons, deuterons, tritons and $^3$He in Au+Au collisions at energies from 0.4A to 1.5A GeV [20]. IQMD transport calculations could reproduce the experimental results using a soft EOS ($K_{nm}$=190±30 MeV) and momentum-dependent interactions [21]. The flow data are sensitive to the EOS, because the collective flow of nucleons is driven by the pressure gradient in the collision zone. Another EOS-sensitive observable was found to be subthreshold strangeness production. According to microscopic transport calculations, for example $K^+$ mesons are produced in secondary collisions of pions and nucleons in heavy-ion collisions, if the kinetic beam energy is below the $K^+$ production threshold in nucleon-nucleon collisions, which is 1.6 GeV. These multi-step processes are enhanced at high density, and, therefore, are sensitive to the EOS. The KaoS collaboration measured subthreshold K+ production in a very heavy and a light collision system at different beam energies [22]. The kaon data could only be reproduced by RQMD model calculations when assuming a soft EOS and taking into account in-medium effects [23]. The data of both the FOPI and the KaoS experiment support values of the nuclear incompressibility in the order of 200 MeV, corresponding to a soft EOS for nuclear matter densities around twice saturation density. These results rule out the stiff Skyrme-type EOS with $K_{nm}$=380 MeV shown in the left panel of figure 5.

The FOPI and ASY-EOS collaborations at GSI have also studied the symmetry energy at baryon densities above saturation density in Au+Au collisions 400A MeV by measuring the elliptic flow of neutrons and protons. By comparing the FOPI data to the results of UrQMD transport calculations, a value of about $E_{sym}$= 60 ±10 MeV was found for 2 $\rho_0$ [24]. From the ASY-EOS data a value of about $E_{sym}$= 55±5 MeV at 2 $\rho_0$ was extracted [25]. This data point is included in the right panel of figure 5.

The currently running heavy-ion collision experiment at GSI is HADES, which is able to measure both hadrons and dielectrons over a large acceptance. The invariant mass spectra of dileptons measured in heavy-ion collision provides information on the temperature of the fireball, because it is not affected by the collective expansion. The slope of the invariant mass spectrum directly reflects the fireball temperature integrated of the collision history, if contributions from vector meson decays are subtracted. This has been recently demonstrated by the HADES collaboration, which measured the di-electron invariant mass spectrum for central Au+Au collisions at 1.23A GeV, as shown in the left panel of figure 6 [26]. The right panel of figure 6 depicts the spectrum after subtraction of contributions from vector meson decays. The resulting exponential slope of the spectrum corresponds to a temperature of about 72 MeV. It is worthwhile to note, that a fireball with a temperature of about 70 MeV and a density of about 2.5 $\rho_0$, which is expected to be reached at this beam energy, has similar properties as the matter in the collision zone of two neutron stars (see figure 4), except for the isospin ratio.

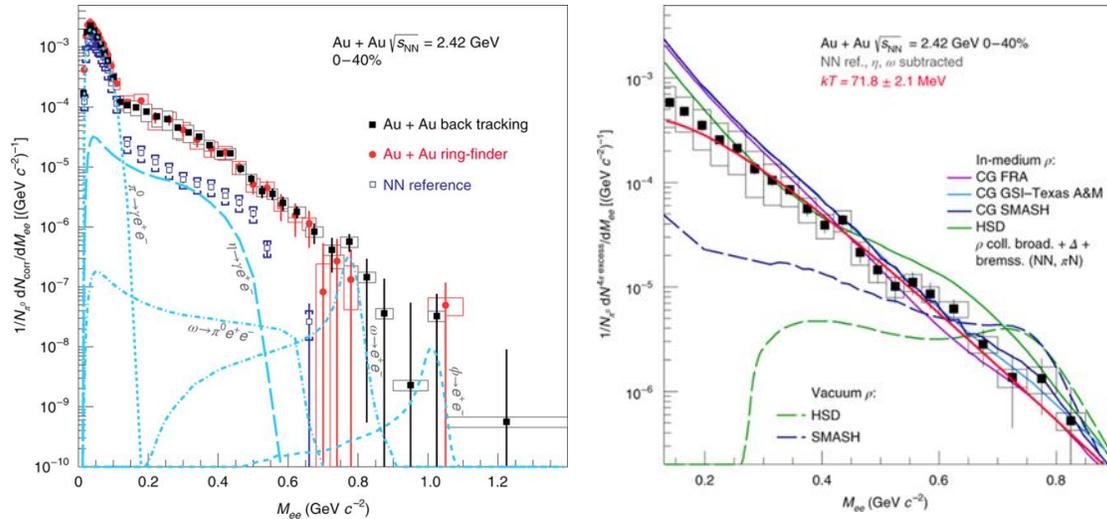

Fig.6: Efficiency corrected dilepton invariant mass spectra measured in central Au+Au collisions at 1.23A GeV [26]. Left: together with a simulated cocktail of contributions from first chance collisions and the freeze-out stage. Right: Contributions from known vector meson decays subtracted. The lines indicate results of different model calculations (see [26]).

Starting in the 1990ies, experiments with gold beams were performed at the AGS in Brookhaven at beam energies between 2A and 11A GeV, i.e. at higher nuclear matter densities. The excitation function of the proton collective flow measured by the EOS collaboration [27] has been analyzed using transport calculations with respect to the high-density EOS [28]. The result is shown in figure 6 as pressure versus density for symmetric nuclear matter. The result of the analysis of the AGS data is represented by the shaded area, which, unfortunately, covers a wide range of nuclear incompressibility, ruling out only very soft or extremely hard EOS. The reason is, that the directed flow data could be explained better by a soft EOS ($K_{nm}$= 210 MeV), while the data on the elliptic flow pointed towards a stiff EOS ($K_{nm}$= 300 MeV). In addition to the AGS data, the results of the analysis of the GSI flow and kaon data are included, shown as the red area below densities of 2 $\rho_0$.

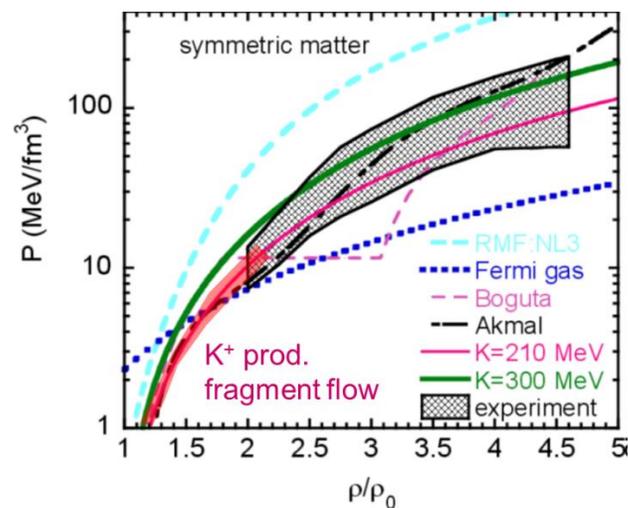

Fig. 6: Zero-temperature EOS for symmetric nuclear matter. The shaded region corresponds to the region of pressures consistent with the experimental flow data. The various curves and lines show predictions for different symmetric matter EOS (see [28]). The red area indicate the results of flow data [21] and kaon data [22] taken at GSI.

### 1.2.2 The quest for the Quark-Gluon Plasma in heavy-ion collisions

Starting 1994, a broad experimental program was launched at the CERN-SPS, devoted to the investigation of high-energy heavy-ion collisions up to beam energies of 160A GeV. In particular, the NA35/NA45 experiments produced a wealth of new data on strangeness production, as illustrated by the red data points in figure 7 [29,30]. The peculiar peak ("horn") in the $\Lambda/\pi^-$ ratio around the collision energy of $\sqrt{s_{NN}} \approx 6$ GeV has been interpreted as an indication for the onset of deconfinement [30]. The NA50/NA60 collaborations investigated the production of charmonium via its dimuon decay in heavy-ion collisions, and found an anomalous suppression of the J/$\psi$ yield in central collisions [31]. This effect was predicted as consequence of the dissociation of charmonium in the QGP [32]. The ratio of the expected to the measured J/$\psi$ yield for central In+In and Pb+Pb collisions at 158A GeV is shown in figure 8.

In the year 2000, CERN announced "compelling evidence" for the discovery of a new state of matter, the Quark-Gluon Plasma (QGP), based on the measurements performed at the SPS, which confirmed key QGP predictions: (i) strangeness enhancement and chemical equilibrium due to shorter time scale for s-sbar production, (ii) charmonium suppression due to colour screening, and (iii) thermal electromagnetic radiation.

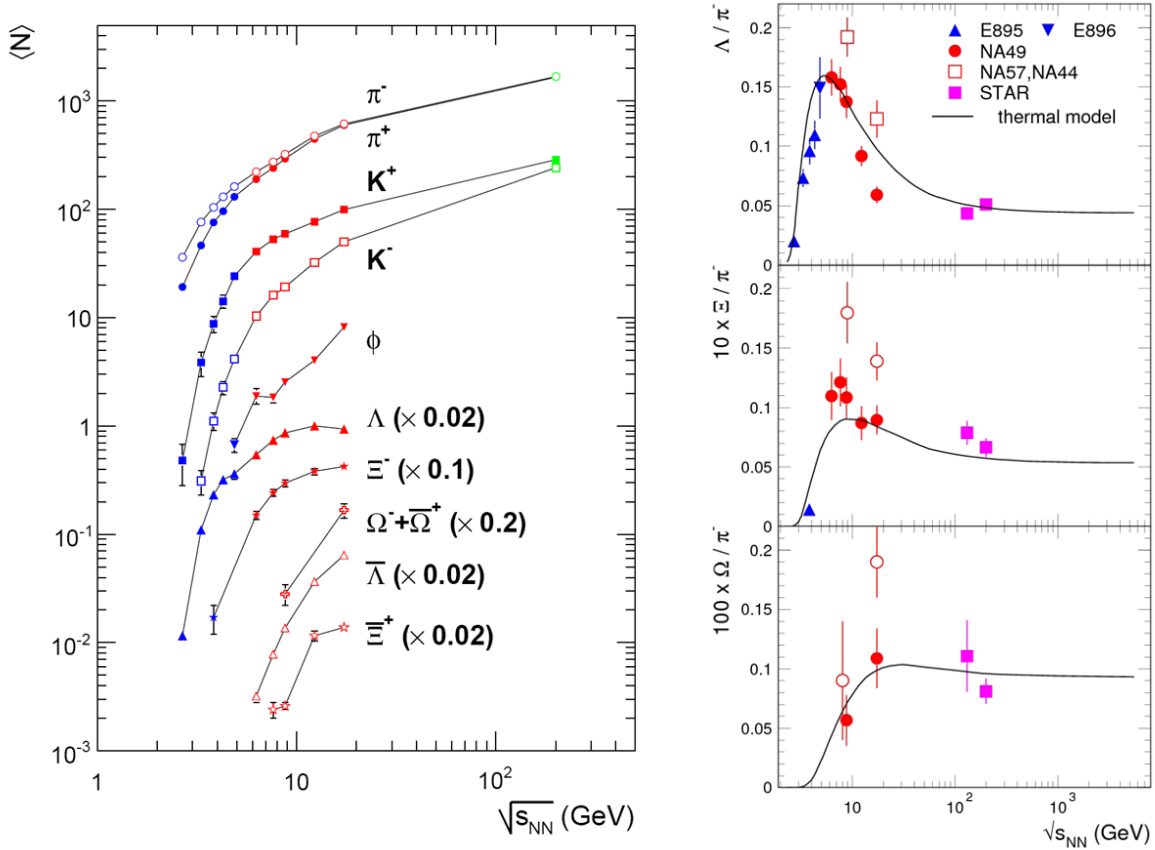

Fig. 7: Particle yields per event (left) and hyperon to pion ratios (right) versus collision energy, measured around midrapidity in central Pb+Pb and Au+Au collisions [29,30].

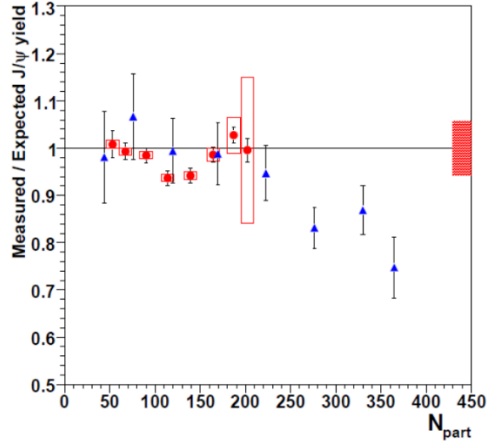

Fig.8: J/ψ suppression pattern in In-In (red) and Pb-Pb (blue) collisions at a beam energy of 158A GeV as function of the number of participants [31].

The NA60 experiment at the CERN-SPS also investigated in detail the dimuon radiation from the fireball at low invariant dimuon masses. Figure 9 depicts the efficiency corrected dimuon excess yield in min. bias In+In collisions at 158 A GeV [33]. All contributions from known dimuon sources, such as the decay of the η, ω, and φ vector mesons including their Dalitz decay, have been subtracted from this spectrum. The spectrum contains only non-trivial contributions as indicated by the coloured lines, such as thermal rates with in-medium ρ and ω spectral functions and free 4π annihilation in hadronic matter, as well as quark-antiquark annihilation in the QGP, folded over a thermal fireball expansion, supplemented with free ρ-meson decays after thermal freeze-out [34]. The figure demonstrates, that a precise measurement of dilepton yields together with a realistic model provides information on medium modifications of hadrons, on the integrated temperature of the fireball, and on the contribution by the QGP.

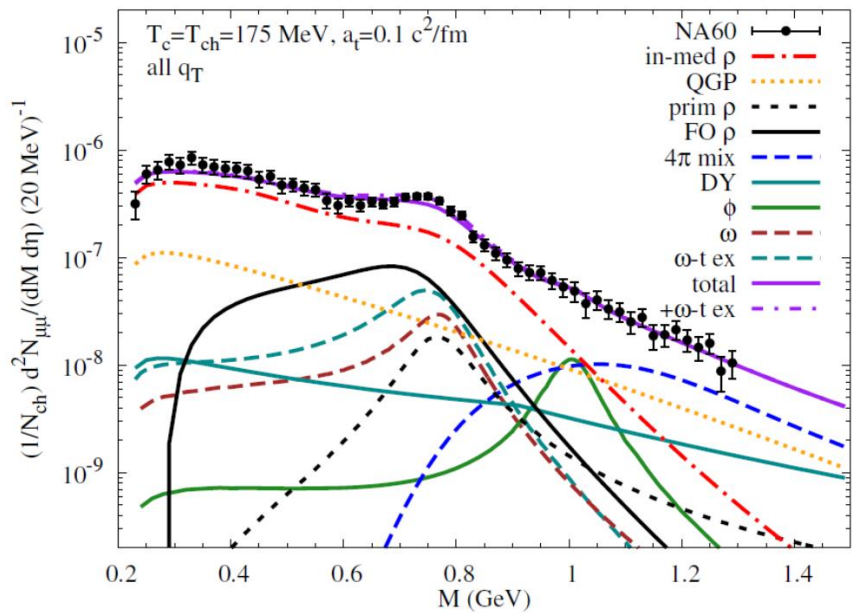

Fig.9: Acceptance corrected dimuon invariant mass spectra measured in In+In collisions at 158 A GeV [33] together with theoretical calculations with in-medium vector spectral functions [34].

## 1.2.3 Searching for the first order phase transition and its critical endpoint

About 10 years ago the STAR collaboration at RHIC/BNL started a beam energy scan program with a distinct goal: the search for the first order phase transition from the QGP to hadron matter and its critical endpoint. The strategy was to identify the collision energy where the QGP signatures observed at the highest RHIC energies turn off, and to find an increase of critical fluctuations of conserved quantities, corresponding to critical opalescence in normal matter. One of the important observables in this respect the elliptic flow of identified particles, which is sensitive to the pressure and the stiffness of the EOS at the very early collision stage. It came as a surprise, that the $v_2$ flow coefficient increases by 50% from to SPS energy ($\sqrt{s_{NN}}$= 17.2 GeV) to top RHIC energy ($\sqrt{s_{NN}}$= 200 GeV).This observation was one of the important indications for the perfect liquid bulk matter dynamics. Moreover, it was found that the elliptic flow of different hadrons follows a universal scaling, when dividing $v_2$ by the number of participant quarks $n_q$ of each hadron, and plotting it as function of the transverse kinetic energy per quark $(m_T-m_0)n_q$. This observation was taken as an argument, that the elliptic flow was generated in the early QGP phase of the collision.

In order to search for the beam energy, at which the universal $v_2$ scaling behaviour disappears, and, hence, hadronization sets in, the STAR collaboration measured the elliptic flow from high energies down to energies, where the collision rate decreased to a few Hz. The result of the elliptic flow beam energy scan is shown in figure 10, which depicts $v_2/n_q$ versus $(m_T-m_0)/n_q$ measured in Au+Au collisions from $\sqrt{s_{NN}}$= 62.4 GeV down to $\sqrt{s_{NN}}$= 7.7 GeV [35]. Scaling of $v_2$ with $n_q$ is observed for all energies and al particles except for the φ meson at the two lowest energies. From the data, no distinct violation of the $n_q$ scaling of $v_2$ can be deduced down to $\sqrt{s_{NN}}$= 7.7 GeV. The antiparticles exhibit a similar $n_q$ scaling, but increasingly deviate in absolute $v_2$ values from the particles with decreasing energy, as illustrated in figure 11. This deviation was attributed to effects of the mean-field potential in both the partonic and the hadronic phase [36]. On the other hand, it was argued that the baryon chemical potential is the determining factor for the observed particle type–dependent splitting in $v_2$ [37]. In conclusion, the search for the disappearance of QGP signals could not be finalized due to the limitations in luminosity of the RHIC collider towards low collision energies.

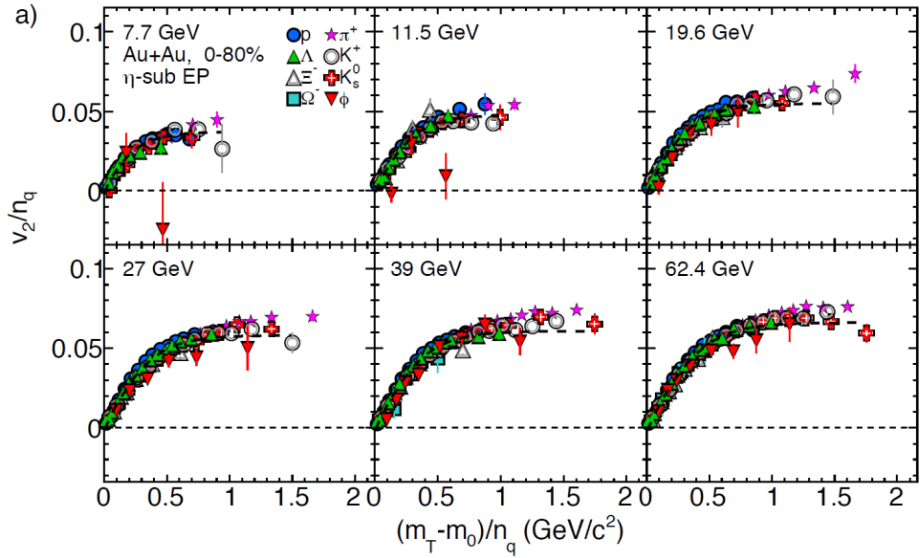

Fig. 10: Elliptic flow $v_2$ scaled by the number of constituent quarks $n_q$ for identified hadrons versus the difference of transverse mass and rest mass scaled by $n_q$ for (0-80%) Au+Au collisions at collision energies from $\sqrt{s_{NN}}$ = 7.7 GeV to 62.4 GeV [35].

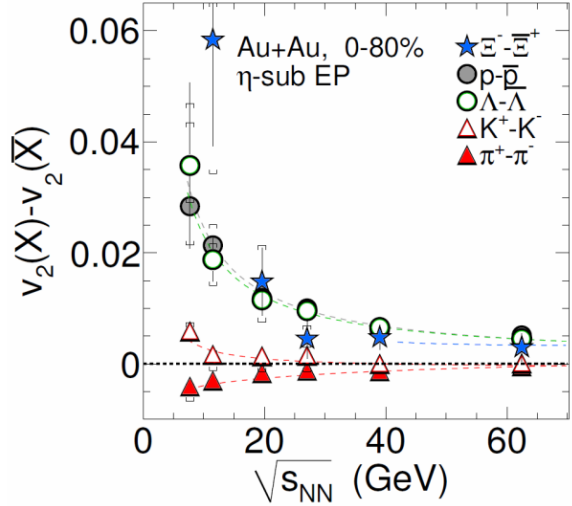

Figure 11: The difference in elliptic flow between particles and their corresponding antiparticles (see legend) as a function of $\sqrt{s_{NN}}$ for 0 - 80% central Au+Au collisions as measured by the STAR collaboration at RHIC [35].

The study of fluctuations in event-by-event multiplicity distributions of conserved quantities, such as baryon number, strangeness, and charge, provides insights into the properties of matter created in high-energy nuclear collisions. Lattice QCD calculations suggest that higher moments of these quantities are sensitive to the phase structure of the hot and dense nuclear matter created in such collisions. Non-Gaussian moments (cumulants) of these fluctuations are expected to be sensitive to the proximity of the critical point. Recent results from the STAR collaboration are shown in figure 12, which depicts the volume-independent cumulant ratio $\kappa\sigma^2$ (excess kurtosis times squared standard deviation) of the net-proton multiplicity distribution as a function of the collision energy, measured in Au+Au collisions [38]. In the absence of a critical point, this quantity should be constant as a function of collision energy, whereas a non-monotonic behaviour of the $\kappa\sigma^2$ observable would signal the location of a critical point. According to figure 12, the STAR-BES data exhibit a deviation from unity for the most central collisions at the lowest measured energy, as expected for a critical behaviour. These results clearly call for a high-precision measurement of higher-order fluctuations at lower beam energies in order to search for the peak in $\kappa\sigma^2$.

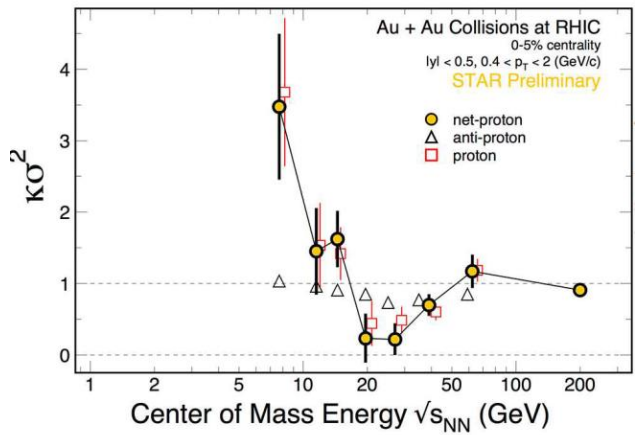

Fig.12: Energy dependence of the product $\kappa\sigma^2$ (excess kurtosis times variance) of the net-proton multiplicity distribution (yellow circles) for top 0-5% central Au+Au collisions. The Poisson expectation is denoted as dotted line at $\kappa\sigma^2=1$ [38].

## 2. Studies of high-density QCD matter at NICA

As reviewed in the previous section, our understanding of the properties of QCD matter at neutron star core densities is still very limited. Future scientific progress in this field will be mainly driven by new experimental results, as lattice QCD is not yet able to make firm predictions for matter at high net-baryon density. In particular, the energy range of NICA, where the highest net-baryon densities are expected to be created, is very poorly explored experimentally: only very few multi-strange (anti-) hyperons have been measured, no hypernuclei have been found, no collective flow data are available except for one proton measurement, no data on event-by-event fluctuations exists, no dilepton spectra have been measured, no charm data are available. Already the systematic and precise measurements of these almost unknown observables have a huge discovery potential, and will contribute to the answers to the following fundamental questions:

- What is the high-density equation-of-state of nuclear matter, which is relevant for our understanding of supernova, the structure of neutron stars, and the dynamics of neutron star mergers?
- What are the relevant degrees of freedom at high densities? Is there a phase transition from hadronic to quark-gluon matter, a region of phase coexistence, and a critical point? Do exotic QCD phases like quarkyonic matter exist?
- Can we find experimental evidence for the restoration of chiral symmetry, in order to shed light on the generation of hadron masses?
- How far can we extend the chart of nuclei towards the third (strange) dimension by producing single and double hypernuclei? Which role do hyperons play in the core of neutron stars?

Heavy-ion collisions at NICA beam energies are ideally suited to provide high net-baryon densities, and to reach a deconfined state of matter. This is illustrated in fig. 13, where the excitation energy density in the center of the collision zone is shown as a function of the net-baryon density for central Au+Au collisions at beam energies from 10A GeV to 40A GeV ($\sqrt{s_{NN}}$ = 4.8 - 8.9 GeV) as predicted by several transport models and a hydrodynamic calculation [39].

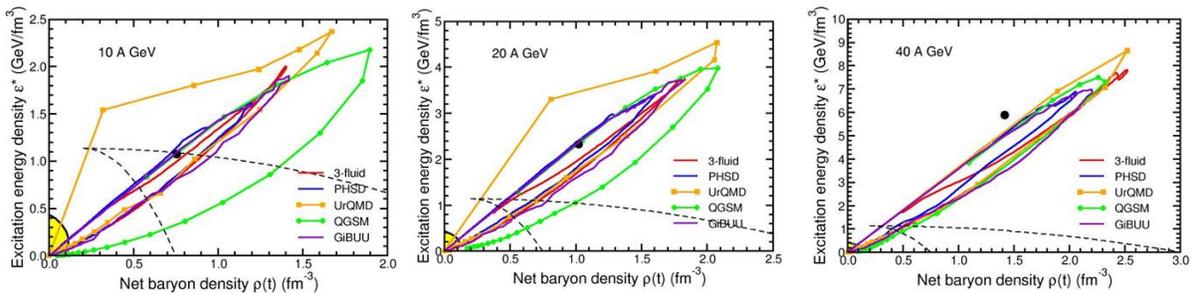

Fig. 13: Phase trajectories ($\rho(t)$, $\varepsilon^*(t)$) for central Au+Au collisions at different bombarding energies calculated by various transport codes and a hydro model [39]. The excitation energy density is defined as as $\varepsilon^*(t) = \varepsilon(t) - m_N\rho$ (see text). The dashed lines indicate the regions of phase coexistence.

The excitation energy is defined as $\epsilon_*(t) = \epsilon(t) - m_N\rho(t)$ with $\epsilon(t)$ the energy density and $m_N\rho(t)$ the mass density. The solid lines correspond to the time evolution of the system; they turn in a clockwise sense, and the dots on the curves labelled UrQMD and QGSM correspond to steps of 1 fm/c in collision time. The dashed lines enclose the expected region of phase coexistence. According to these model calculations, both the energy density and the net-baryon density are sufficiently high in order to produce a plasma of quarks and gluons.

The NICA complex fills a gap in the energy landscape of existing and future accelerator facilities as illustrated in figure 14 [40]. The NICA collider is designed to run at a maximum luminosity of L=$10^{27}$ cm$^{-2}$s$^{-1}$ at collision energies between $\sqrt{s_{NN}}$ = 8 and 11 GeV corresponding to a reaction rate of 6 kHz for minimum bias Au+Au collisions, exceeding the available rates at STAR/RHIC and at NA61/CERN-SPS by about two orders of magnitude in this energy range. At energies below $\sqrt{s_{NN}}$ = 8 GeV, the collision rate at NICA luminosity decreases to about 100 Hz at $\sqrt{s_{NN}}$ = 5 GeV. The CBM detector at FAIR is a fixed target experiment and is designed to run at extremely high interaction rates up to 10 MHz for selected observables such as J/ψ, and at 1-5 MHz for multi-strange hyperons and dileptons.

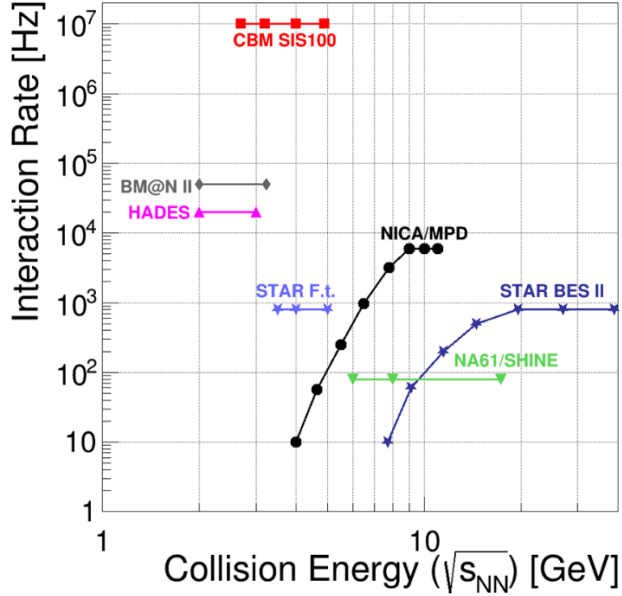

Fig. 14: Interaction rates achieved by existing and heavy-ion experiments under construction as a function of the center-of-mass energy [40].

### 2.1 Promising observables at MPD/NICA

In order to characterize the properties of the fireball matter one has to measure the abundantly produced particles including their phase-space distributions, correlations and fluctuations, together with rare diagnostic probes like multi-strange particles, lepton pairs and charmed hadrons. The experimental challenge will be to identify these rare particles with high purity, and to collect samples with large statistics, which is required for a multi-differential analysis.
In the following, we discuss the most promising observables in the NICA energy range, which are expected to be sensitive to the high-density equation-of-state and to possible phase transitions, and, hence, will help to understand the structure of neutron stars and the dynamics of neutron star mergers.

#### 2.1.1 Probing the high-density EOS by the collective flow of identified particles

As mentioned above, the collective flow of identified particles is driven by the pressure gradient in the reaction volume, and, hence, is sensitive to the nuclear matter-equation-of state. In case the EOS softens by a phase transition, also the collective flow might be modified. This effect is illustrated in figure 15, which shows the elliptic flow of identified particles, integrated over transverse momentum, as calculated for Au+Au collisions at different energies [41]. The calculation was performed using a three-fluid hydrodynamics model employing three different EOS: a purely hadronic EOS, one with a first-order phase transition, and one with a crossover transition. According to the model, the elliptic flow for particles is very similar for the different EOS, whereas the flow of antiparticles is modified by the phase transition.

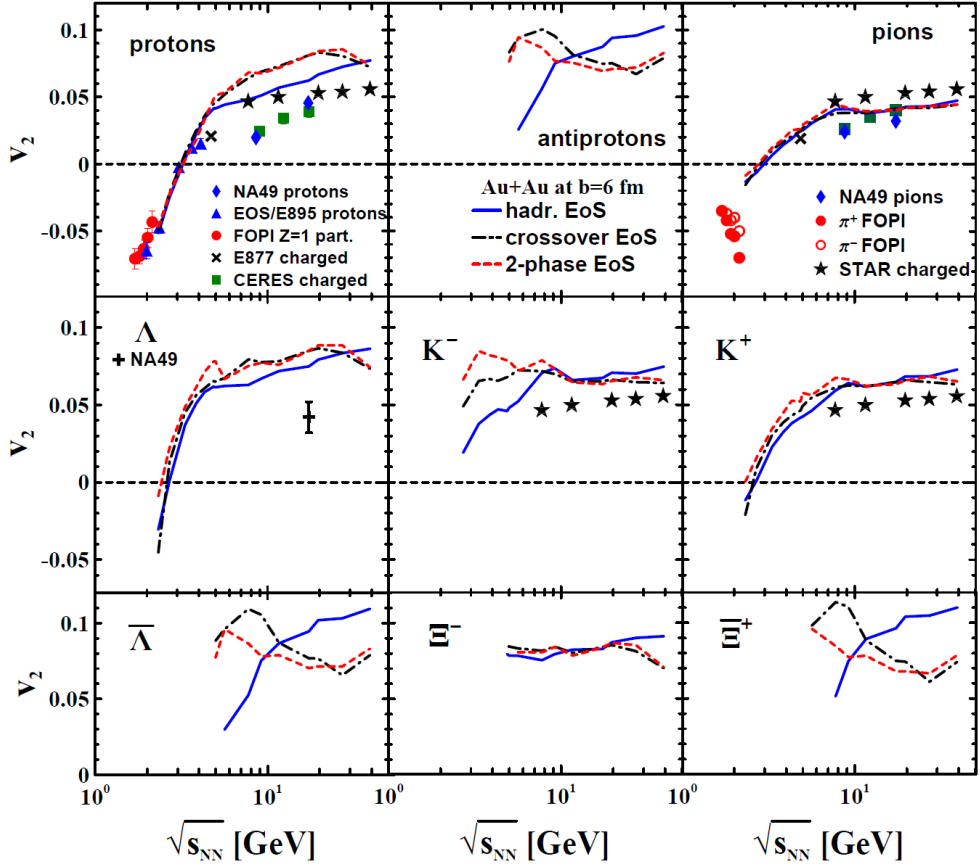

Fig. 15: Transverse-momentum integrated (0 <$p_t$< 2 GeV/c) elliptic flow of various hadrons at midrapidity as a function of incident energy in mid-central collisions Au+Au at b = 6 fm [41].

In the NICA energy range, no flow data exist for anti-baryons, except for anti-protons and anti-lambda at $\sqrt{s_{NN}}$ = 7.7 GeV measured by the STAR collaboration (see figure 11). Moreover, in the AGS energy range only the proton flow has been measured [27]. At these energies, the flow of particles like Ω hyperons or φ mesons would be of particular interest, because they do not suffer from rescattering because of their low hadronic cross sections.

In conclusion, the precise determination of the elliptic flow of particles and anti-particles (including (anti-) Ω hyperons) across the NICA energy range would dramatically improve the data situation. These measurements are expected to shed light on the high-density EOS, and, possibly, on the existence of a phase transition at NICA energies.

## 2.1.2 Strangeness: exploring the EOS and searching for the onset of deconfinement

According to microscopic transport models, multi-strange (anti-)hyperons are predominantly produced in heavy-ion collision via multiple steps involving kaons and lambdas at beam energies close to or even below the production threshold. These sequential collisions happen more frequently at high densities, and, therefore, the yield of multi-strange (anti-) hyperons increases with increasing density of the fireball. The sensitivity of multi-strange hyperon production on the density of the reaction volume, and, hence, on the EOS, has been calculated with the novel PHQMD transport mode [42]. Preliminary results indicate, that in Au-Au collisions at lower NICA energies (<$\sqrt{s_{NN}}$ = 5 GeV), where baryon densities of up to 8 $\rho_0$ are reached, the expected yields of $\Xi^-$ and $\Omega^-$ hyperons and even more of $\Xi^+$ and $\Omega^+$ anti-hyperons, are substantially higher for a soft EOS than for a stiff EOS.

The statistical hadronization model describes well the measured yields of hadrons and even of light (anti-) nuclei produced in ultra-relativistic heavy-ion collisions, supporting the scenario of a chemically and thermally equilibrated fireball, which freezes-out at a temperature of about 156 MeV [43]. The fact, that also the yields of $\Omega^-$ and $\Omega^+$ hyperons agree with the model assumptions, although the omega-nucleon scattering cross section is small, was interpreted as a signature for a phase transition from the quark-gluon plasma to a hadronic final state, which drives multi-strange hyperons into equilibration [44]. The yields of multi-strange hyperons measured in Pb + Pb collisions at a beam kinetic energy of 30 A GeV at the CERN-SPS were also found to be in agreement with the statistical model, as illustrated in the left panel of figure 16 [45]. In measurements of a smaller collision system (Ar+KCl) at a much lower beam energy (1.76 A GeV), the yield of $\Xi^-$ hyperons was found to exceed the statistical model prediction by about a factor 24 ± 9, although the abundantly produced hadrons agree with the thermal model for a temperature of about T = 76 MeV, as illustrated in the right panel of figure 16 [46]. This finding indicates a non-equilibrium production mechanism of $\Xi^-$ hyperons.

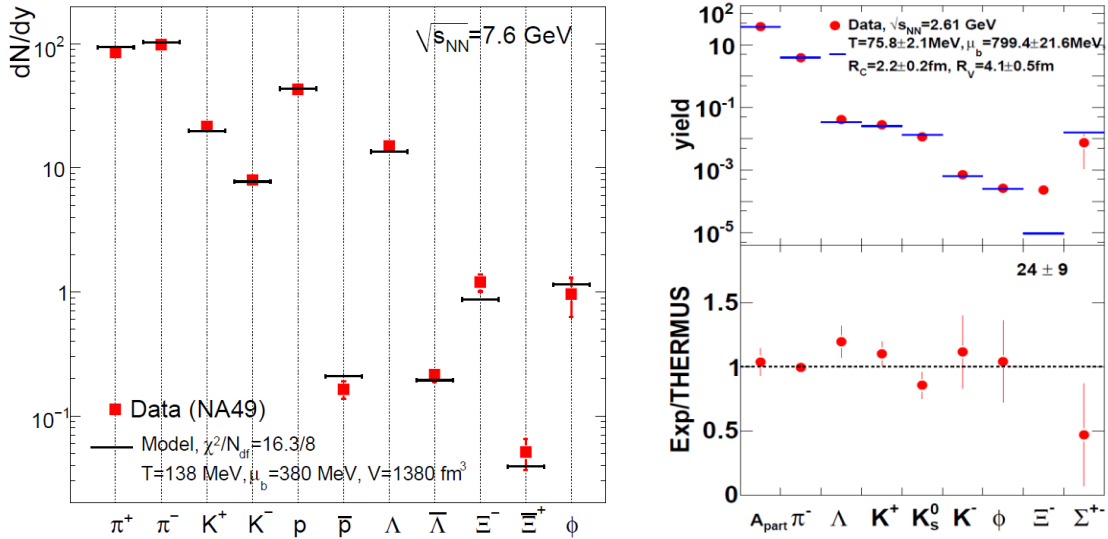

Fig.16: Left: Experimental hadron yields measured by the NA49 collaboration in Pb+Pb collisions at 30A GeV compared to a thermal model fit (taken from [45]). Right upper panel: Yields of secondary hadrons measured in Ar+KCl reactions at 1.76A GeV by the HADES collaboration and the corresponding THERMUS fit (blue bars) [46]. The lower plot shows the ratio of the experimental value and the SHM value. For the $\Xi^-$ hyperon the ratio number is quoted instead of a point.

In conclusion, the precise measurement of the excitation function of the yields of multi-strange (anti-) hyperons in the NICA energy range will provide information on the high-density EOS, and on the energy, where these particles reach thermal equilibrium, which is regarded as an indication for the onset of deconfinement.

### 2.1.3 Scouting the critical point by event-by-event fluctuations of conserved quantities

As discussed above, the STAR collaboration found an increase of the 4$^{th}$ order cumulant (kurtosis) of the net-proton multiplicity distribution towards the lowest collision energy of $\sqrt{s_{NN}}$ = 7.7 GeV. Such an effect was expected to occur if the particles freeze-out in the vicinity of the critical point, similar to the phenomenon of critical opalescence, indicating a first order phase transition. However, in order to discover the maximum of the fluctuation, and, hence, the possible location of the QCD critical endpoint, the measurement has to be extended towards lower collision energies. This is hardly possible at RHIC because of the steeply decreasing luminosity of the collider towards lower energy. Moreover, recent lattice QCD calculations suggest, that a possible critical endpoint at physical values of the quark masses and for non-vanishing baryon-chemical potential can occur only at a temperature $T_{cep}$ < 140 MeV [47]. Such a temperature corresponds to rather low collision energies, which do not allow precision experiments with STAR in the RHIC collider mode.

In conclusion, as suggested by experimental indications and theoretical considerations, the NICA accelerator covers the energy range, where the QCD critical endpoint is supposed to be located, and, hence, the MPD detector has the potential to discover it - or to rule it out.

### 2.1.4 The quest for the chiral phase transition by dilepton spectroscopy

The precise multi-differential measurement of lepton pairs in heavy-ion collision provides a wealth of information, for example on the in-medium modification of vector mesons, the temperature of the fireball, chiral symmetry restoration and phase transitions. The most precise dilepton measurement up-to-date has been performed with muon pairs produced in In+In collisions at 158 A GeV by the NA60 collaboration [33]. The resulting dilepton excess (i.e. after subtraction of known dilepton sources) at low invariant masses is shown in figure 9 together with model calculations [34]. According to the calculations, various processes contribute to the measured yield including a broadened in-medium ρ meson, radiation from the QGP, and dileptons from 4 π mixing which can be related to ρ-$a_1$ chiral mixing. Up to date, no dilepton data have been taken in heavy-ion collisions in the NICA energy range.

The left panel in figure 17 depicts a dielectron invariant mass spectrum simulated for central Au+Au collisions at 20A GeV ($\sqrt{s_{NN}}$= 6.45 GeV) [40]. The solid red curve shows the contribution of the thermal radiation which includes in-medium ρ, ω, 4-π spectral functions and QGP spectrum calculated using the many-body approach of [48]. This thermal radiation dominates the invariant mass spectrum in the range between 1 and 2 GeV/$c^2$. It is worthwhile to note that the spectral slope is not blue-shifted, i.e. not affected by the radial expansion of the fireball, and, hence, is directly related to the temperature of the fireball, integrated over the collision history. This offers the unique opportunity, to experimentally determine the caloric curve of QCD matter, when measuring the dilepton invariant mass spectrum at different collision energies, i.e. the excitation function of the average temperature of the emitting source. This temperature is plotted in the right panel of figure 17 for two cases: The red dashed line depicts the temperature derived from the intermediate dilepton invariant mass range (1–2 GeV/$c^2$), calculated with a fireball model and a coarse-graining approach [49]. Such a curve reflects a smooth crossover transition, or no phase transition. The purple solid line illustrates a hypothetical caloric curve. Shown are also the temperature measurements of HADES [26] and NA60 [50]. The experimental discovery of the caloric curve would proof phase coexistence, and provide indications for the onset of deconfinement and the location of the critical endpoint. The flow of lepton pairs as function of their invariant mass would allow to disentangle radiation from the early partonic phase and the late hadronic phase.

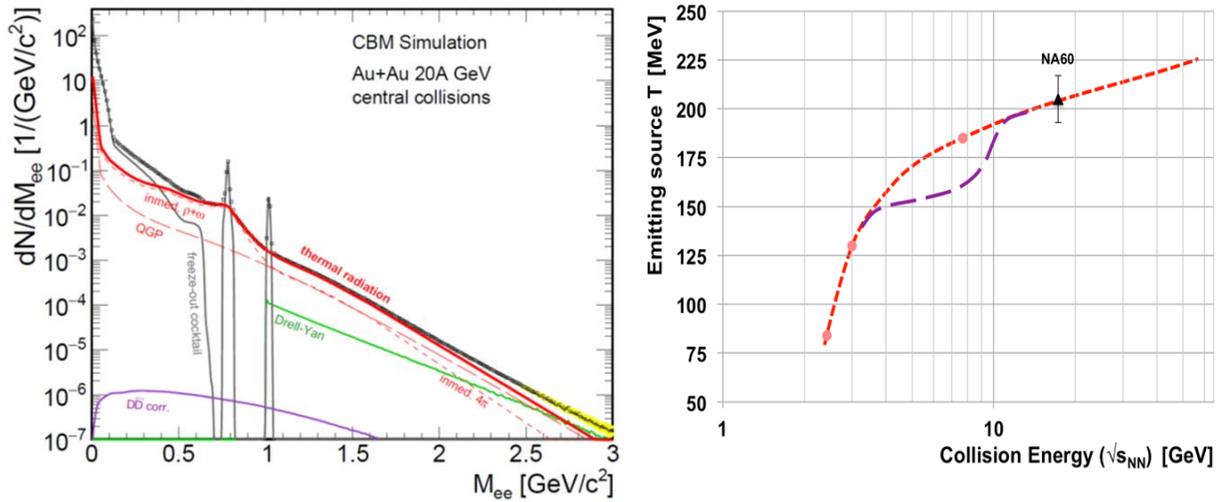

Fig. 17: Left panel: Invariant-mass spectrum of e⁺e⁻ pairs radiated from a central Au+Au collision at 20AGeV(see text) [40]. Right panel: Sketch of the emitting source temperature as function of collision energy for two scenarios [40]. The red dashed line illustrates the temperature calculated with a fireball model and a coarse-graining approach, extracted from the dilepton invariant mass range between 1–2 GeV/c² [49]. Such a curve is expected for a crossover transition, or for purely hadronic matter. (see text). The purple line illustrates a hypothetical caloric curve reflecting a first order phase transition. The black dots show experiments result from HADES [26] and NA60 [50].

The experimental challenge of dilepton measurements is the very low signal cross section and the high combinatorial background. A decent measurement of dileptons at invariant masses around the rho-meson or between 1 - 2 GeV/c² requires a signal-to-background ratio of at least S/B = 1/100. In this case, one needs about 10000 signal pairs in order to determine the yield with a statistical accuracy of 10%. Moreover, the systematic error due to the subtraction of the combinatorial background should be in the order of $10^{-3}$. According to these estimates, for dilepton measurements NICA should run near the maximum luminosity, i.e. above a collision energy of about $\sqrt{s_{NN}}$ = 8 GeV. At these energies, also exploratory measurements of charmonium production could be performed at NICA.

### 2.1.5 The charming probe of the QGP

Particles containing charm quarks are expected to be created in the very first stage of the reaction, and, therefore, offer the possibility to probe partonic degrees-of- freedom. Depending on their interaction with the medium, the charm and antticharm quarks hadronize into D mesons, charmed baryons, or charmonium. As discussed above, the suppression of charmonium due to color screening of the heavy quark potential in the deconfined phase is the first predicted signature for quark-gluon plasma formation, and has been found experimentally at the SPS. In heavy-collisions at LHC energies, however, charmonium is less suppressed. This enhancement with respect to SPS energies is interpreted as a signal for deconfined, thermalized charm quarks, which hadronize at the phase boundary, and (re-)generate charmed hadrons [51].
The world data set on the total charm production cross section is shown in figure 18 [52]. No data on open and hidden charm production in heavy-ion collisions are available at beam energies below 158A GeV. Pioneering measurements of open and hidden charm production at NICA would shed light on the charm production mechanism at beam energies around $\sqrt{s_{NN}}$ = 10 GeV and below. Moreover, the yield ratio of charmonium to open charm measured in heavy-ion collisions at different beam energy could provide information on the degrees-of freedom in the fireball. This is illustrated in figure 19, which depicts the ratio of J/ψ mesons to D + anti-D mesons calculated for Au+Au collisions in the NICA energy range. The ratio is calculated using two different model, i.e. the Handron String Dynamics (HSD) code [53], which describes the production of charmed particles in a hadronic environment, and the Statistical Hadronization model (SHM) [54], where charm and anti-charm quarks are created in a deconfined phase. The HSD and SHM results are represented by the red and blue symbols, respectively. The hadron model predicts the ratio of J/ψ to (D + anti-D) mesons to increase with decreasing collision energy. This effect is due to the different thresholds for charm production in nucleon-nucleon collisions, which is $\sqrt{s_{NN}}$ = 4.97 GeV for J/ψ, $\sqrt{s_{NN}}$ = 5.1 GeV for anti-D+$\Lambda_c$, and $\sqrt{s_{NN}}$ = 5.61 GeV for D+anti-D. Therefore, the yield of J/ψ increases with respect to the D + anti-D meson yield for collision energies approaching the charmonium production threshold. In the SHM calculation, c and anti-c quarks are assumed to be produced in the deconfined phase, and the hadrons are created at freeze-out. Hence, no individual production thresholds exist, and the SHM model predicts a more or less energy-independent ratio. The measurement of J/ψ mesons, $\Lambda_c$ hyperons and D mesons in Au+Au collisions at NICA would certainly help to unravel the question, in which state of matter they are produced in.

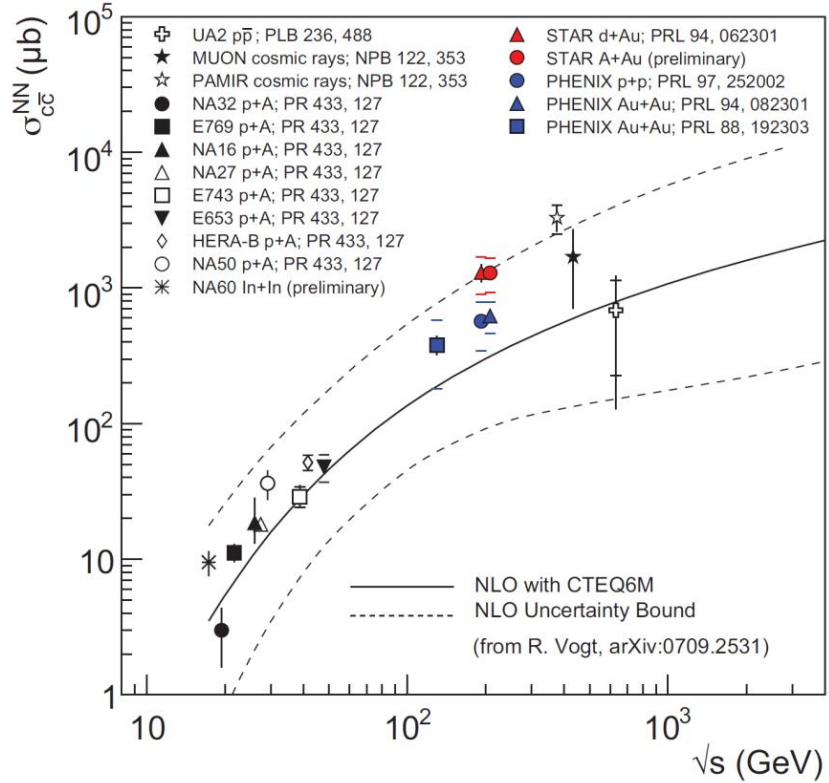

Fig.18: Total charm production cross section measurements compared to NLO calculations with uncertainty bound. Taken from [52].

Information on the charm production mechanism can also be extracted from the absolute yields of charmonium and D mesons measured in heavy-ion collisions. Recent calculations based on the UrQMD event generator predict charm production in multiple-step processes, which might happen in a hadronic scenario [55]. The UrQMD code includes known N* resonances with given widths, which are excited in sequential nucleon-nucleon collisions, and then decay with a given branching ratio into charmed particles, like N* → $\Lambda_c$ + anti-D and N* → N +J/ψ. This effect increases the yield of J/ψ mesons and anti-D mesons by several orders of magnitude around the production threshold, and there is even a substantial yield below threshold. The measurement of such an unexpected high charm yield clearly would suggest, that these particles are created in a hadronic scenario.

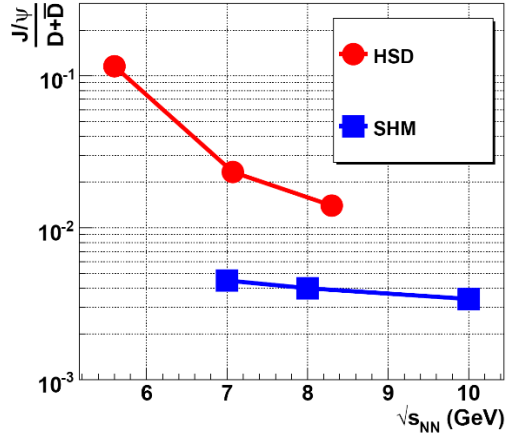

Fig. 19: Ratio of J/$\psi$ mesons to D + anti-D mesons calculated for Au+Au collisions as function of collision energy with the Hadron-String-Dynamics (HSD) transport code (red symbols) [53], and with the Statistical Hadronization Model (SHM) (blue symbols) [54]. While the HSD model describes a hadronic scenario, SHM assumes the production of c c-bar pairs in a deconfined phase.

**2.1.6 Hypernuclei and the hyperon puzzle in neutron stars**

As discussed above, the high-density EOS is the key to our understanding of neutron stars. To accommodate masses of about two solar masses or more, the EOS should be sufficiently stiff. Any phase transition would open new degrees-of-freedom, and, hence, soften the EOS. In order to allow for quark matter in neutron star cores, the softening of the EOS should be prevented by effects like repulsive vector potentials between the quarks [9]. However, the high-density EOS might not only be softened by a phase transition from hadronic to quark matter, but also by hyperons, which should appear at high densities, if the chemical potential of neutrons and protons exceeds the chemical potential of hyperons. In order to solve this "hyperon puzzle," various models propose different mechanisms to circumvent a softening of the EOS at high densities. The results of calculations based on chiral effective field theory assuming only hadronic degrees-of-freedom are illustrated in figure 20, which depicts the chemical potentials of neutrons and lambdas as function of density [56]. The model features a repulsive $\Lambda$N potential together with repulsive 3-body $\Lambda$NN interactions at high densities, and, thus, prevents the condensation of lambdas below densities of about 5 $\rho_0$.

The $\Lambda$N, $\Lambda$NN and $\Lambda\Lambda$ interaction is also responsible for the stability of (double-) lambda hypernuclei. Therefore, these interactions can be investigated by identifying new hypernuclei, and by measuring their lifetime. The NICA energy range is perfectly suited to study the production of hypernuclei in heavy-ion collisions. According to calculations based on a statistical model, the yield of light (double-) lambda hypernuclei exhibits a maximum in the FAIR energy range, as shown in figure 21 [57]. This maximum is due to the fact, that the hyperon yield increases with increasing beam energy, while the yield of light nuclei decreases with increasing beam energy. Measurements at NICA offer the opportunity to improve our knowledge on the lifetimes of known light hypernuclei, and to discover new ones, including double-lambda hypernuclei.

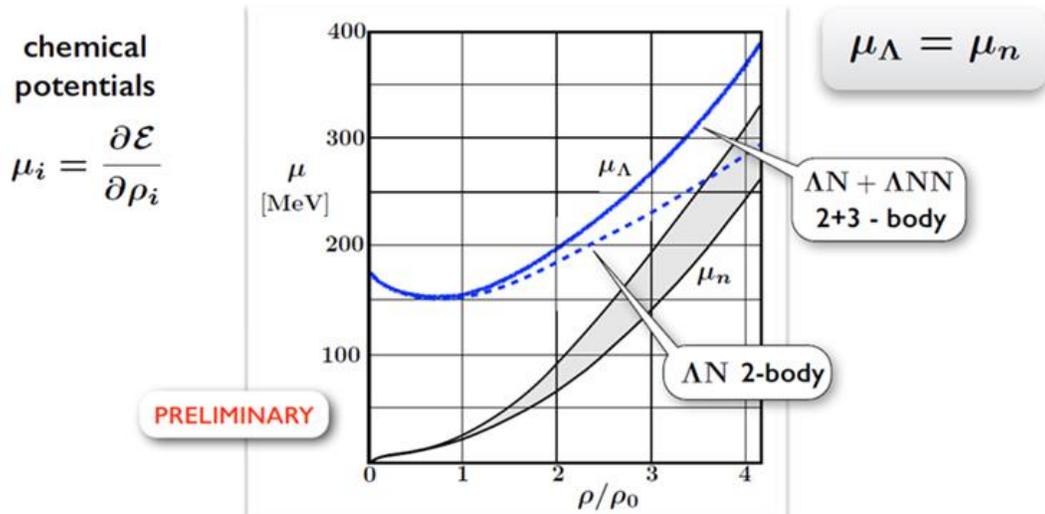

Fig. 20: Chemical potentials of Λ hyperon $\mu_\Lambda$ and neutron $\mu_n$ in neutron matter as a function of baryon density. The lambda chemical potential is calculated taking into account ΛN two-body interactions only (blue-dashed line) and including ΛNN three-body forces (blue solid line). The grey-shaded area between the black lines indicates the neutron chemical potential [56].

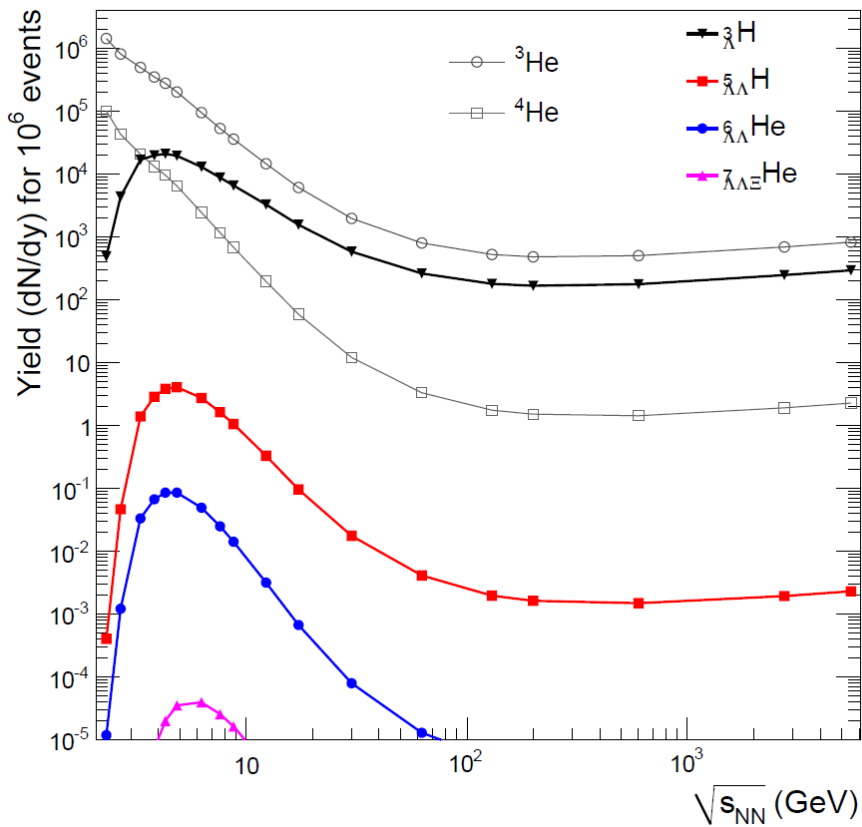

Figure 21: Energy dependence of hypernuclei yields at midrapidity for $10^6$ central collisions as calculated with a thermal model. The predicted yields of $^3$He and $^4$He nuclei are included for comparison [57].

### 3. Conclusions and outlook

The beam energies and luminosities available at NICA perfectly fit into the worldwide landscape of heavy-ion accelerators. Situated between FAIR SIS100 and the low RHIC energies, NICA covers the energy range where the highest net-baryon densities can be created in laboratory experiments. This offers the opportunity to produce and to study QCD matter at neutron star core densities, and to address fundamental questions related to the high-density EOS, to the phase structure of strongly interacting matter, to confinement and chiral symmetry.

As a modern heavy-ion experiment, the Multi-Purpose Detector (MPD) at NICA [58] is designed to measure a variety of diagnostic probes as discussed in the previous sections. This feature allows the investigation of different observables, which address the same physics topic. For example, the high density EOS can be studied by (i) the elliptic flow of identified particles, and (ii) the production of multi-strange (anti-hyperons). A deconfined phase should leave its traces in several observables, including (i) the collective flow, (ii) the high invariant mass slope of dileptons, and (iii) charm production. The order of the phase transition can be determined by the (i) caloric curve determined with dileptons, the (ii) critical point measured with event-by-event fluctuations, and by (iii) the onset of deconfinement extracted from the excitation function of multi-strange hyperons.

The successful execution of these measurements with MDP at NICA will achieve a breakthrough in our understanding of the properties of QCD matter under extreme conditions. This is particularly true, if the different results form a consistent picture, for example, if the caloric curve obtained from dilepton measurements ends at the same energy, where event-by-event fluctuations exhibit a maximum, indicating the critical endpoint. Such a concordant scenario would tremendously increase the persuasive power of the single observations, and could be regarded as the experimental discovery of a first order phase transition in dense baryonic matter. The experiments with MPD at NICA will complement the observations made by the beam-energy scan of STAR, and will be complemented towards lower energies by the CBM experiment at FAIR. At even lower energies, the BM@N experiment at NICA will contribute to the progress in the field, although with a reduced set of observables.

The experimental program at MPD requires high-statistics measurements of the relevant observable over a large phase space, with particle identification and event characterization capabilities. The core detector of the MPD is a Time-Projection-Chamber (TPC) inside a solenoid magnet providing tracking and particle identification via dE/dx measurements. Particle identification will be improved by a Time-of-Flight (TOP) detector based on Multigap Resistive Plate Chambers (MPRC). These two detector systems, together with an Electromagnetic Calorimeter (ECAL), will also provide electron identification. A Zero Degree Calorimeter (ZDC) will be used to determine the collision centrality and the orientation of the reaction plane.

The TPC will be complemented by an Inner Tracking System (ITS) in order to provide precise tracking, momentum determination and vertex reconstruction. This upgrade will improve the identification of hyperons via the topology of their weak decays ($\Lambda \to p\,\pi$, $\Xi \to \Lambda\,\pi$, $\Omega \to \Lambda\,K$). Due to the short decay lengths of $c\tau = 7.89$ cm ($\Lambda$), $c\tau = 4.91$ cm ($\Xi^-$), and $c\tau = 2.46$ cm ($\Omega^-$) the track measurement should start close to the primary vertex, where the track density is high. In particular, the ITS will be required for the identification of D mesons via their hadronic decay into pions and kaons (decay length of $c\tau = 123$ µm for $D^0$, and $c\tau = 312$ µm for $D^\pm$), and of and $\Lambda_c^+$ hyperons via their decay into $pK^-\pi^+$ with a decay length of only 60 µm. The layout, the technical realization, and the performance of the ITS is described in the Technical Design Report.

## Acknowledgment

The author acknowledges support from the Europeans Union's Horizon 2020 research and innovation programme under grant agreement No. 871072, and from the Ministry of Science and Higher Education of the Russian Federation, grant N 3.3380.2017/4.6 and by the National Research Nuclear University MEPhI in the framework of the Russian Academic Excellence Project (contract No. 02.a03.21.0005, 27.08.2013).
## References

[1] NUPECC Long Range Plan 2017
[2] S. Borsanyi et al., JHEP 1009 (2010) 073
[3] A. Basavov et al., Phys. Rev. D 85 (2012) 054503
[4] F. Becattini et al., Phys. Rev. Lett. 111 (2013) 082302
[5] A. Andronic et al., Nature 2018, 561, 321
[6] K. Fukushima and T. Hatsuda, Rept. Prog. Phys. 74 (2011) 014001
[7] L. McLerran L and R. Pisarski, Nucl. Phys. A 796 (2007) 83
[8] G. Baym et al., Rep. Prog. Phys. 81 (2018) 056902
[9] M. Orsaria et al., Phys. Rev. C89 (2014)015806
[10] E. Most E et al., Phys. Rev. Lett. (2019) 122 061101
[11] T. Klaehn et al., Phys. Rev. C 74 (2006) 035802
[12] H. T. Cromartie et al. 2020 Nature Astronomy 4 72
[13] C. Fuchs, PoSCPOD07:060 (2007)
[14] J.P. Blaisot et al., 1995, Nucl. Phys. A 591 (1995) 435
[15] B.-A. Li and X. Han, Phys. Lett. B 727 (2013) 276
[16] T. E. Riley et al., Astrophys. J. Lett. 887 (2019) L21.
[17] C. D. Capano et al., arXiv 2019, arXiv:1908.10352v2.
[18] H.H. Gutbrod, A.M. Poskanzer and H.G, Ritter, Rep. Prog. Phys. 52,10
[19] J. W. Harris et al., Phys. Lett. B 153 (1985) 377-381
[20] A. Le Fèvre et al., Nucl. Phys. A945, (2016) 112–133.
[21] A. Le Fevre A et al., (The FOPI Collaboration) Nucl. Phys. A 945 (2016) 112
[22] C. Sturm et al., (The KaoS Collaboration) Phys. Rev. Lett. 86 (2001) 39.
[23] C. Fuchs et al., Phys. Rev. Lett. 86 (2001) 1974
[24] Y. Leifels et al., (The FOPI Collaboration) 1993 Phys. Rev. Lett. 71 963
[25] P. Russotto et al., (The ASY-EOS Collaboration) 2016 Phys. Rev. C 94 034608
[26] J. Adamczewski-Musch et al. (The HADES Collaboration) Nature Physics 15 (2019) 1040
[27] C. Pinkenburg et al, Phys. Rev. Lett. 83 (1999) 1295.
[28] P. Danielewicz, R. Lacey, and W.G. Lynch, W.G. Science 298 (2002) 1592.
[29] C. Blume et al. (NA49 Collaboration) nucl-ex/0409008
[30] C. Alt et al., (NA49 Collaboration) Phys.Rev.C77 (2008) 024903
[31] R. Arnaldi et al., (Na60 Collaboration) Nucl.Phys.A830 (2009) 345c
[32] T. Matsui and H. Satz, Phys. Lett. 178 (1986) 416
[33] R. Arnaldi et al., (Na60 Collaboration) Eur. Phys. J. C 61, (2009) 711
[34] R. Rapp, J. Wambach and H. van Hees, in Relativistic Heavy-Ion Physics, edited by R. Stock, Landolt Börnstein (Springer), New Series I/23A (2010) arXiv:0901.3289 hep-ph
[35] L. Adamczyk et al. (STAR Collaboration) Phys. Rev. C88 (2013) 014902
[36] J. Xu et al., Phys. Rev. Lett. 112 (2014) 012301
[37] Y. Hatta, A. Monnai, B.W. Xiao, Nucl. Phys. A 947(2016) 155